\documentclass[referee,epjc3]{svjour3}

\usepackage{epsfig}
\usepackage{graphicx}
\usepackage{amsfonts,amssymb}
\usepackage{color}

  \usepackage{slashed}
  \usepackage{url}
\usepackage{bm}
\usepackage{verbatim}
\usepackage{float}

\def\beq{\begin{equation}}
\def\eeq{\end{equation}}
\def\bea{\begin{eqnarray}}
\def\eea{\end{eqnarray}}
\def\ba{\begin{array}}
\def\ea{\end{array}}

\journalname{Eur. Phys. J. C}
\begin{document}


\title{Running mass of the gluon propagator in the Landau gauge}


\author{Marco Frasca\thanksref{e1,addr1}}
\thankstext{e1}{e-mail: marcofrasca@mclink.it}
\institute{Via Erasmo Gattamelata, 3 \\
             00176 Roma (Italy) \label{addr1}}


\date{\today}

\maketitle

\begin{abstract}
Gluon propagator in the infrared limit, computed in the Landau gauge, can be cast into an universal form with a running mass for the gluon. In this way we are able to show that all the proposals appeared so far in literature are equivalent and describe the same physics. In this way, it appears essential to reach a general agreement about the interpretation of such results. We discuss the points that make difficult to agree on a similar view about this.
\end{abstract}




\section{Introduction}

Since the initial proposal by Gribov and Zwanzinger \cite{Gribov:1977wm,Zwanziger:1989mf}, the study of correlation functions, mostly the propagators, of the Yang-Mills theory in the Landau gauge has seen a lot of activity both theoretical and through lattice computations. Gribov and Zwanzinger yielded arguments favoring a gluon propagator running to zero and a ghost propagator running to infinity faster than a free one when momenta go to zero, the infrared limit. This proposal was crucial to understand confinement as, being this the state of art of our understanding, all the criteria for a confined Yang-Mills theory were promptly satisfied. Indeed, this proposal entails a non trivial infrared fixed point for a pure Yang-Mills theory in absence of quarks. Few years later, this area of research has known a significant growing interest thanks to the work by Alkofer, von Smekal and Hauck \cite{vonSmekal:1997is,vonSmekal:1997vx}, that used a truncated scheme to solve Dyson-Schwinger equations and by Cornwall \cite{Cornwall:1981zr}, that introduced the so-called pinch technique in perturbation computations \cite{Cornwall:2010bk}. But while the work by Alkofer, von Smekal and Hauck supported the conclusions by Gribov and Zwanzinger, Cornwall's approach obtained a significant different result: Gluon propagator did not seem to go zero rather it reached a finite non-null value in the infrared limit. At that time, lattice computations were not able enough to decide what was the right behavior. On the other side, some other theoretical works supported Cornwall's conclusions \cite{Aguilar:2004sw,Boucaud:2005ce,Boucaud:2011ug,Frasca:2007uz} following different approaches.

Finally, on 2007, due to the improvement of computational resources, it was possible to reach volumes large enough to verify the behavior of propagators in the Landau gauge at lower momenta doing lattice computations. The results confirmed that the gluon propagator reaches a finite non-null value at lower momenta and the ghost propagator behaves like that of a free massless particle \cite{Bogolubsky:2007ud,Cucchieri:2007md,Oliveira:2007px}. Thereafter, the original proposal by Gribov and Zwanzinger was properly corrected to fit lattice data \cite{Dudal:2008sp} as also confirmed from lattice computations \cite{Dudal:2010tf}. Today, this result has been widely accepted for three and four dimensions while for the two-dimensional case the original scenario by Gribov and Zwanzinger,
dubbed scaling solution,
applies \cite{Maas:2007uv}.
This means that the scaling solution is not meaningful for our analysis but could have a meaning in three and four dimensions in QCD when quarks are present as recently shown in \cite{Hopfer:2014zna}.

Our aim in this paper is to show how all the theoretical approaches presented in literature so far to analyze propagators in Landau gauge for Yang-Mills theory without quarks yield identical results in the deep infrared. This means that, whatever approach one likes, it is possible to do any kind of computations with the theories devised so far. In order to achieve this conclusion, we just use the original idea of Cornwall \cite{Cornwall:1981zr} that, in the infrared limit, the gluon propagator is characterized by a running mass that reaches a constant value at very small values of momenta. This idea applies straightforwardly to whatever approach was conceived so far to discuss Yang-Mills theory in the infrared limit. The only critical point is to understand the meaning of such a result in terms of running coupling. There are diverging views about this but the facts that perturbation theory applies successfully in the infrared limit, as shown by Tissier and Wschebor \cite{Tissier:2010ts,Tissier:2011ey}, and renormalization group arguments confirm the triviality of the theory in the infrared limit, as shown by Weber \cite{Weber:2011nw}, seem to seriously hint that the running coupling should go to zero lowering momenta. This was seen on the lattice \cite{Bogolubsky:2009dc} with the definition of the running coupling given therein.

The paper is so structured. In Sec.~\ref{sec2} we introduce the gluon propagator in the Landau gauge and show how the different approaches yield it in closed analytical form. In Sec.~\ref{sec3} we give a definition of running mass and evaluate it for all the approaches we discuss. The equivalence is seen at numerical level between them being perfectly coincident below 1~GeV. In Sec.~\ref{sec4} we discuss the running coupling and yields a triviality argument depending on the preceding analysis of the running mass. Finally, in Sec.~\ref{sec5} conclusions are presented.

\section{Gluon propagator}
\label{sec2}

In the Landau gauge, the gluon propagator takes the generic form
\begin{equation}
    \Delta_{ab}^{\mu\nu}(p)=\delta_{ab}\left(\eta^{\mu\nu}-\frac{p^\mu p^\nu}{p^2}\right)\Delta(p)
\end{equation}
being $a,b$ color indexes and $\eta^{\mu\nu}$ the metric tensor. In the proposals seen in current literature, the gluon propagator in the Landau gauge can be generally represented through the equation
\begin{equation}
   \Delta^{-1}(p)=Z^{-1}\left[p^2+M^2(p)\right]
\end{equation}
where we assume an Euclidean signature, $M(p)$ is a running mass and $Z$ a constant factor. The running mass has the property to reach a finite value when $p\rightarrow 0$, in agreement with lattice data, and becomes negligibly small in the opposite limit \cite{Oliveira:2010xc}. This formula yields also a definition of a running mass by inversion as
\begin{equation}
\label{eq:RM}
    M^2(p)=-p^2+Z\Delta^{-1}(p)
\end{equation}
that can be immediately used in numerical calculations. The evaluation of the running mass is meaningful just below about 1 GeV where the coupling is large enough that perturbation theory cannot be trusted anymore. In all the studies given in literature the constant $Z$ can be taken to be unity. The running mass will depend on the gauge choice as it depends on the gluon propagator that changes with the given gauge.

The current proposals presented in literature, by different approaches, can be resumed as follows. The propagator proposed by Cornwall \cite{Cornwall:1981zr} and then refined in a series of works by Papavassiliou, Aguilar and Binosi can be cast in the form \cite{Aguilar:2010gm,Bicudo:2015rma}
\begin{equation}
    \Delta^{-1}(p)=m^2+p^2\left[1+\frac{13Ng^2}{96\pi^2}\ln\left(\frac{p^2+\rho m^2}{\mu^2}\right)\right]
\end{equation}
where $m$, $g$ and $\rho$ are taken as free fitting parameters and $\mu$ is the renormalization point taken to be 3 GeV. From this form of the propagator the derivation of the running mass is straightforward and is given by
\begin{equation}
\label{eq:CPAB}
    M^2_{CPAB}(p)=m^2+p^2\frac{13Ng^2}{96\pi^2}\ln\left(\frac{p^2+\rho m^2}{\mu^2}\right).
\end{equation}
We assume this formula to hold up to $p\approx 1\ GeV$. This expression holds in the Landau gauge and will change with a different choice of the gauge \cite{Bicudo:2015rma}.

In the infrared limit, Tissier and Wschebor used perturbation theory after inserting a mass term for the gluon in the Lagrangian. Their propagator takes the form \cite{Tissier:2010ts,Tissier:2011ey}
\begin{eqnarray}
    \Delta^{-1}_{TW}(p)&=&p^2+m^2+\frac{p^2Ng^2}{384\pi^2}\left[111\frac{m^2}{p^2}
		-2\frac{m^4}{p^4}+\left(2-\frac{p^4}{m^4}\right)\ln\left(\frac{p^2}{m^2}\right)+\right. \nonumber \\
		&&2\left(\frac{m^2}{p^2}+1\right)^3
		\left(\frac{p^4}{m^4}-10\frac{p^2}{m^2}+1\right)\ln\left(1+\frac{p^2}{m^2}\right)+ \nonumber \\
&&\left(4\frac{m^2}{p^2}+1\right)^\frac{3}{2}
\left(\frac{p^4}{m^4}-20\frac{p^2}{m^2}+12\right)\ln\left(\frac{\sqrt{4+\frac{p^2}{m^2}}
-\sqrt{\frac{p^2}{m^2}}}{\sqrt{4+\frac{p^2}{m^2}}+\sqrt{\frac{p^2}{m^2}}}\right)- \nonumber \\
&&\left.\left(p^2\rightarrow\mu^2\right)\right].
\end{eqnarray}
and the running mass can be immediately obtained as
\begin{eqnarray}
\label{eq:TW}
    M^2_{TW}(p)&=&m^2+\frac{p^2Ng^2}{384\pi^2}\left[111\frac{m^2}{p^2}
		-2\frac{m^4}{p^4}+\left(2-\frac{p^4}{m^4}\right)\ln\left(\frac{p^2}{m^2}\right)+\right. \nonumber \\
		&&2\left(\frac{m^2}{p^2}+1\right)^3
		\left(\frac{p^4}{m^4}-10\frac{p^2}{m^2}+1\right)\ln\left(1+\frac{p^2}{m^2}\right)+ \nonumber \\
&&\left(4\frac{m^2}{p^2}+1\right)^\frac{3}{2}
\left(\frac{p^4}{m^4}-20\frac{p^2}{m^2}+12\right)\ln\left(\frac{\sqrt{4+\frac{p^2}{m^2}}
-\sqrt{\frac{p^2}{m^2}}}{\sqrt{4+\frac{p^2}{m^2}}+\sqrt{\frac{p^2}{m^2}}}\right)- \nonumber \\
&&\left.\left(p^2\rightarrow\mu^2\right)\right].
\end{eqnarray}
This equation was obtained from perturbation theory and, as such, could be improved by higher order corrections. The parameter $m$ is the gluon mass introduced initially in the Lagrangian and should be fixed to fit while $\mu$ is the renormalization point.

\begin{figure}[H]
\begin{center}
\includegraphics[angle=0, width=\textwidth]{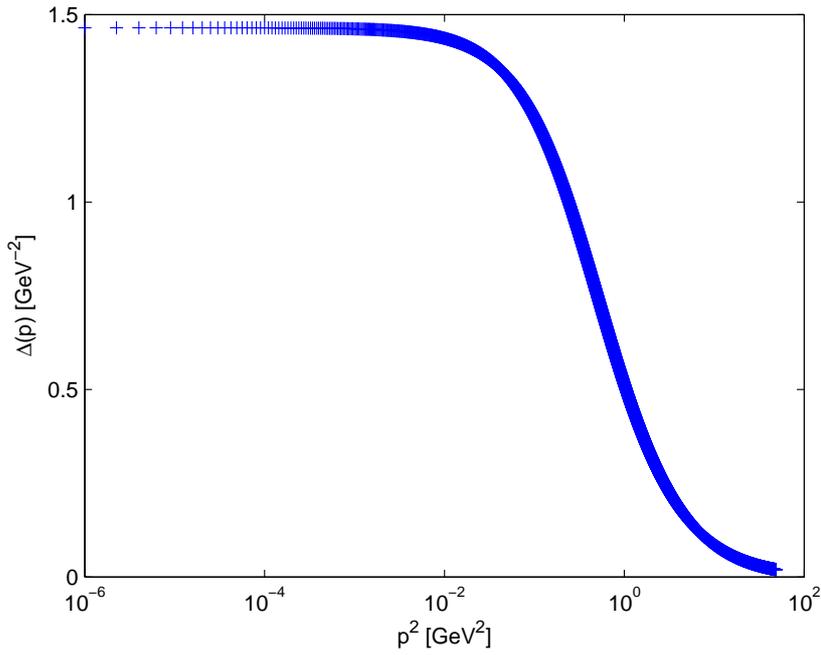}
\caption{\label{fig:fig1} Numerical evaluation of the gluon propagator in the Landau gauge as given in eq.(\ref{eq:mine}), for $m_0=0.7\ GeV$, in close agreement with lattice computations.}
\end{center}
\end{figure}

Postulating the existence of a dimension-two condensate, Dudal, Gracey, Sorella, Vandersickel and Verschelde proposed a gluon propagator in the form \cite{Dudal:2008sp}
\begin{equation}
\label{eq:DGSVV}
   \Delta_{DGSVV}(p)=\frac{p^2+M_0^2}{p^4+(M_0^2+m_0^2)p^2+2Ng^2\gamma^4+m_0^2M_0^2}
\end{equation}
being $m_0$, $M_0$ and $\gamma$ free fitting parameters. A study has been accomplished comparing this form with lattice data \cite{Cucchieri:2011ig} and the best fit was obtained for
\begin{equation}
\label{eq:DGSVV2}
   \Delta_{DGSVV}(p)=\frac{\alpha_+}{p^2+\omega_+^2}+\frac{\alpha_-}{p^2+\omega_-^2}
\end{equation}
that can be easily rewritten in to the form given in eq.(\ref{eq:DGSVV}). But this is also a very good approximation to the propagator we proposed some time ago \cite{Frasca:2007uz} and recently refined \cite{Frasca:2013kka} in the form
\begin{equation}
\label{eq:mine}
   \Delta_F(p)=\sum_{n=0}^\infty\frac{B_n}{p^2+m_n^2}
\end{equation}
where we put
\begin{equation}
\label{eq:Bn}
    B_n=(2n+1)^2\frac{\pi^3}{4K^3(-1)}\frac{e^{-(n+\frac{1}{2})\pi}}{1+e^{-(2n+1)\pi}}.
\end{equation}
and $m_n=(2n+1)\frac{\pi}{2K(-1)}m_0$. Here $m_0$ is a free fitting parameter and $K(-1)$ is the complete elliptic integral of the first kind. It is easy to see that each term in (\ref{eq:mine}) is exponentially damped and so, the propagator in eq.(\ref{eq:DGSVV2}) is an excellent approximation to it. In the following we will assume that the parameters of eq.(\ref{eq:DGSVV2}) are always chosen to best fit eq.(\ref{eq:mine}). The reason for doing so is that, based on the preceding observation, these two forms of propagator should give the same results when used to fit lattice data and the running mass is almost the same in the momenta region we aim to study. In any case, the running mass can be evaluated using eq.(\ref{eq:RM}) also noticing that
\begin{equation}
    \sum_{n=0}^\infty B_n=1.
\end{equation}
This propagator is represented in Fig.~\ref{fig:fig1} and is similar to that observed on lattice computations that also all other proposals excellently fit.

\section{Running mass}
\label{sec3}

\begin{figure}[H]
\begin{center}
\includegraphics[angle=0, width=\textwidth]{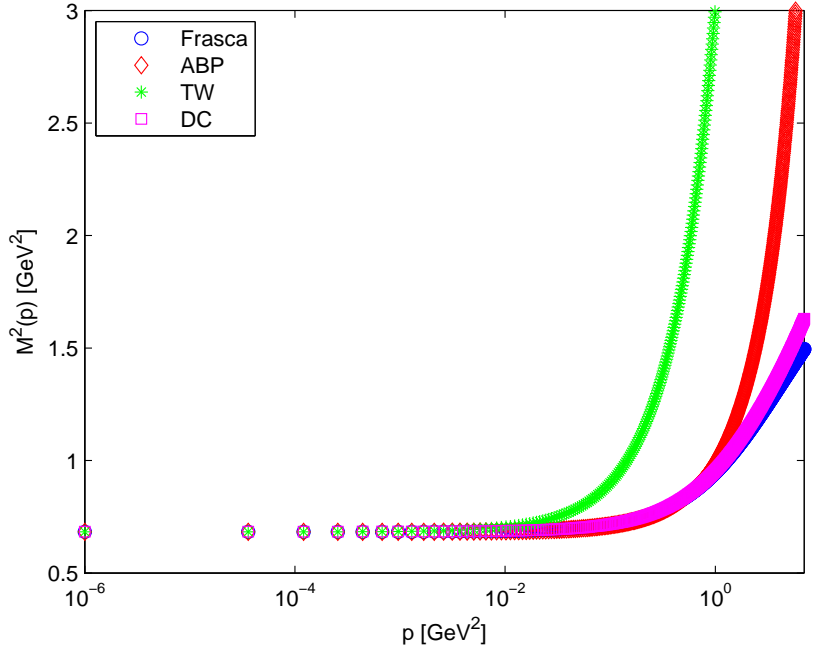}
\caption{\label{fig:fig2} Numerical evaluation of the running masses for all the approaches discussed in Sec.~\ref{sec2} (see text for further details).}
\end{center}
\end{figure}

\begin{figure}[H]
\begin{center}
\includegraphics[angle=0, width=\textwidth]{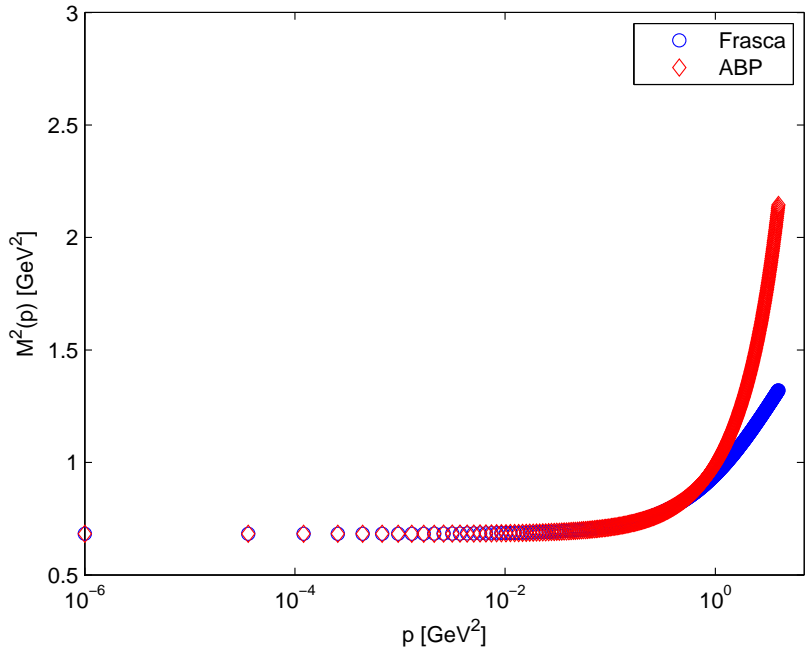}
\caption{\label{fig:fig3} Numerical evaluation of the running masses just for CPAB and from propagator (\ref{eq:mine}).}
\end{center}
\end{figure}

When evaluated, running masses for all the devised approaches yield the cumulative behavior displayed in Fig.~\ref{fig:fig2}. Here TW means Tissier and Wschebor, CPAB means Cornwall, Papavassiliou, Aguilar and Binosi and DC means Dudal and Cucchieri referring to the numerical fits in \cite{Cucchieri:2011ig} for the propagator of Dudal, Gracey, Sorella, Vandersickel and Verschelde.
In Fig.~\ref{fig:fig3} we show just the fit with the CPAB running mass of the running mass given by eq.(\ref{eq:mine}) and it is clearly evident the excellent agreement in the region below 1 GeV.
These fits are provided with the parameter $m$ in eq.(\ref{eq:TW}) and eq.(\ref{eq:CPAB}) to be the same as $m_0$ in eq.(\ref{eq:mine}). In any case we assume for the renormalization point $\mu=3\ GeV$. This applies both to TW and CPAB. For CPAB we also take $\rho=27$ and $g_0^2=9.77$. We assume SU(3) everywhere so that $N=3$. All the running masses agree excellently till momenta around 1 GeV as expected except for TW. Latter discrepancy is due to the fact that TW running mass is obtained by perturbation theory at one loop and so is amenable to improvements going to higher orders. We expect in the end an excellent agreement for all the approaches discussed here. This result has some implications. Firstly, whatever the preferred approach the results are always in agreement with lattice data but in some cases there are more parameters to adjust the fit. This is not relevant in view of the fact that the qualitative picture they give is a correct one. Secondly, the running mass given in CPAB, eq.(\ref{eq:CPAB}), is a really good fit for all the cases we discussed so far and represents an excellent resummation for eq.(\ref{eq:mine}). Thirdly, it appears a fundamental fact that a better interpretation of this equation for the running mass will give a better understanding of Yang-Mills theory in the infrared. The key point here is to reach an agreement about the theory being or not trivial in the infrared limit. Common wisdom assumes that the theory should reach a finite non-null infrared fixed point and the coupling should have a finite non-null coupling at zero momenta. There is no proof whatsoever of such a statement for a pure Yang-Mills theory while this is certainly true for QCD when also quarks are included. The main difficulty to proof such a statement relies on the difficulties on share a common view on how an infrared running coupling should be properly defined. Fourthly, all these approaches draw the important conclusion that Yang-Mills theory has indeed a mass gap but none of them provides a really complete mathematical proof.

\section{Running coupling}
\label{sec4}

There are some important results that add to the conclusion that Yang-Mills theory becomes free in the low-energy limit without quarks. We try to summarize them here. As discussed in the preceding section, Yang-Mills theory with a mass term added reproduces quite well lattice data using standard perturbation theory \cite{Tissier:2010ts,Tissier:2011ey}. Recently, it was shown by Weber and Dall'Olio \cite{Weber:2014lxa} that this perturbation technique can be used to solve Callan-Szymanzik equations for the theory in the infrared. The running coupling is shown to go to zero at lower momenta as already shown by Weber using renormalization group arguments \cite{Weber:2011nw}. This would be in perfect agreement with lattice studies (see \cite{Bogolubsky:2009dc} and Refs. therein) and the rather commonly accepted definition of the running coupling as
\begin{equation}
\label{eq:RC}
    \alpha_s(p)=\frac{g_0^2}{4\pi}Z(p)J^2(p)
\end{equation}
assuming for the gluon propagator $\Delta(p)=Z(p)/p^2$ and for the ghost propagator $G(p)=J(p)/p^2$. It easy to see that, with this definition, all the cases we discussed in Sec.~\ref{sec2} will have the running coupling going to zero as momenta go to zero. This is due to the mass gap that all these approaches share and with the fact that, for the ghost propagator, $J(p)\approx 1$ when $p\rightarrow 0$. It is also fundamental to note that the running mass has the property that
\begin{equation}
   M^2(p)\stackrel{p\rightarrow 0}{\rightarrow}m^2
\end{equation}
and so we are left with a Yukawa propagator proper to a free theory in agreement with triviality requirements. So, the existence of a mass gap implies that the theory would be trivial in the infrared limit.

Finally, let us consider a propagator in the form
\begin{equation}
   \Delta(p)=\sum_{n=0}^\infty\frac{Z_n}{p^2-m_n^2+i\epsilon}
\end{equation}
with $B_n>0$ for any $n$ and $m_n$ the mass spectrum. In this case, K\"allen-Lehman representation can be applied with
\begin{equation}
   \sigma(\mu^2)=\sum_{n=0}^\infty Z_n\delta(\mu^2-m_n^2).
\end{equation}
Given the case that $\sum_{n=0}^\infty Z_n=1$, the theory has no bound states and so it is trivial. This is exactly the case provided in eq.(\ref{eq:mine}) that fits excellently well with the other proposals in the low energy limit in agreement with Ref.~\cite{Frasca:2006yx} for scalar field theory. Indeed, we can write the approximate equation at small momenta, in the Euclidean case and using eqs.(\ref{eq:mine}), (\ref{eq:Bn}), the corresponding spectrum for $m_n$ and eq.(\ref{eq:CPAB}),
\begin{equation}
  -p^2+\left(\sum_{n=0}^\infty\frac{B_n}{p^2+m_n^2}\right)^{-1}\approx m^2+p^2\frac{13Ng^2}{96\pi^2}\ln\left(\frac{p^2+\rho m^2}{\mu^2}\right)
\end{equation}
provided
\begin{equation}
   m^2=\left(\sum_{n=0}^\infty\frac{B_n}{m_n^2}\right)^{-1}.
\end{equation}
Then, the ghost sector decouples from the the Yang-Mills field and a free massless propagator holds in the deep infrared limit. We can evaluate the running coupling through eq.(\ref{eq:RC}) by noticing that $J(p)\approx 1$ and
\begin{equation}
   Z(p)\approx\frac{p^2}{p^2+m^2+p^2\frac{13Ng^2}{96\pi^2}\ln\left(\frac{p^2+\rho m^2}{\mu^2}\right)}
\end{equation}
and then
\begin{equation}
\label{eq:RCeff}
    \alpha_s(p)\approx\frac{g_0^2}{4\pi}\frac{p^2}{p^2+m^2+p^2\frac{13Ng^2}{96\pi^2}\ln\left(\frac{p^2+\rho m^2}{\mu^2}\right)}
\end{equation}
that goes to zero like $p^2$ as seen on lattice \cite{Bogolubsky:2009dc} using the same definition of the running coupling. This fully justifies the successful perturbative approach of Tissier and Wschebor in the infrared limit \cite{Tissier:2010ts,Tissier:2011ey}.

Quantum chromodynamics without quarks does not exist in nature. This means that a Yang-Mills theory could be trivial but adding quarks can change the characteristic of the infrared fixed point. This same thing is seen in the ultraviolet limit where the beta function depends critically on the number of quarks. For the latter case one has
\begin{equation}
   \beta(\alpha_s)=-\left(\frac{11}{3}N-\frac{2N_q}{3}\right)\frac{\alpha_s^2}{2\pi}
\end{equation}
where $N$ is the number of colors and $N_q$ is the number of quarks. We see that asymptotic freedom is preserved in QCD only if $N_q<11N/2$ otherwise quarks could change the characteristics of the theory at higher energies. This is exactly what happens in the infrared limit as has been recently shown in Ref.~\cite{Hopfer:2014zna} by solving Dyson-Schwinger equations.
   
\section{Conclusions}
\label{sec5}

The analysis performed for different gluon propagators proposed in literature so far has shown how these apparently different approaches should be considered equivalent below 1~GeV instead. The idea is that anyone of these could be reformulated through a proper definition of a running mass. This idea was firstly put forward by Cornwall on 1982 together with the view that the gluon propagator should display a mass gap when momenta are small enough. This scenario appears vindicated by lattice computations. Anyhow, it is interesting to note that the idea that a pure Yang-Mills theory should hit a non-trivial fixed point in the infrared limit, without any kind of proof whatsoever, was so rooted that, even if Cornwall obtained a Yukawa propagator in the deep infrared, no argument of triviality was put forward at all. Since then, arguments are mounting about this conclusion and, for some authors, this is no more a taboo. This will be material for sociology of science in the future.

The most important conclusion to be drawn is that we have now a great arsenal of ammo to treat Yang-Mills theory in the infrared limit that is worth to be exploited and applied to more involved problems of quantum chromodynamics. My hope for the future is to see the low-energy spectrum of the theory completely under control.





\end{document}